\documentclass[%
 reprint,
superscriptaddress,
 amsmath,amssymb,
 aps,
]{revtex4-2}

\usepackage{graphicx}
\usepackage{dcolumn}
\usepackage{bm}
\usepackage{float}

\begin{document}

\preprint{APS/123-QED}

\title{Coherent driving of direct and indirect excitons in a quantum dot molecule}

\author{Frederik Bopp}
\email{frederik.bopp@wsi.tum.de}
\affiliation{%
 Walter Schottky Institut, School of Natural Sciences, and MCQST, Technische Universität München, Am Coulombwall 4, 85748 Garching, Germany
}%

\author{Johannes Schall}
\affiliation{%
 Technische Universität Berlin, Hardenbergstraße 36, 10623 Berlin, Germany
}%

\author{Nikolai Bart}
\affiliation{%
 Faculty of Physics and Astronomy, Ruhr-Universität Bochum, Universitätsstraße 150, 44801 Bochum, Germany
}%

\author{Florian Vogl}
\affiliation{%
 Walter Schottky Institut, School of Natural Sciences, and MCQST, Technische Universität München, Am Coulombwall 4, 85748 Garching, Germany
}%

\author{Charlotte Cullip}
\affiliation{%
 Walter Schottky Institut, School of Natural Sciences, and MCQST, Technische Universität München, Am Coulombwall 4, 85748 Garching, Germany
}%

\author{Friedrich Sbresny}
\affiliation{%
 Walter Schottky Institut, School of Computation, Information and Technology, and MCQST, Technische Universität München, Am Coulombwall 4, 85748 Garching, Germany
}%

\author{Katarina Boos}
\affiliation{%
 Walter Schottky Institut, School of Computation, Information and Technology, and MCQST, Technische Universität München, Am Coulombwall 4, 85748 Garching, Germany
}%

\author{Christopher Thalacker}
\affiliation{%
 Walter Schottky Institut, School of Natural Sciences, and MCQST, Technische Universität München, Am Coulombwall 4, 85748 Garching, Germany
}%

\author{Michelle Lienhart}
\affiliation{%
 Walter Schottky Institut, School of Natural Sciences, and MCQST, Technische Universität München, Am Coulombwall 4, 85748 Garching, Germany
}%

\author{Sven Rodt}
\affiliation{%
 Technische Universität Berlin, Hardenbergstraße 36, 10623 Berlin, Germany
}%

\author{Dirk Reuter}
\affiliation{%
Paderborn University, Department of Physics, Warburger Straße 100, 33098 Paderborn, Germany
}%

\author{Arne Ludwig}
\affiliation{%
 Faculty of Physics and Astronomy, Ruhr-Universität Bochum, Universitätsstraße 150, 44801 Bochum, Germany
}%

\author{Andreas Wieck}
\affiliation{%
 Faculty of Physics and Astronomy, Ruhr-Universität Bochum, Universitätsstraße 150, 44801 Bochum, Germany
}%

\author{Stephan Reitzenstein}
\affiliation{%
 Technische Universität Berlin, Hardenbergstraße 36, 10623 Berlin, Germany
}%

\author{Kai Müller}
\affiliation{%
 Walter Schottky Institut, School of Computation, Information and Technology, and MCQST, Technische Universität München, Am Coulombwall 4, 85748 Garching, Germany
}%

\author{Jonathan J. Finley}%
 \email{finley@wsi.tum.de}
\affiliation{%
 Walter Schottky Institut, School of Natural Sciences, and MCQST, Technische Universität München, Am Coulombwall 4, 85748 Garching, Germany
}%

\date{\today}

\begin{abstract}
Quantum dot molecules (QDMs) are one of the few quantum light sources that promise deterministic generation of one- and two-dimensional photonic graph states. The proposed protocols rely on coherent excitation of the tunnel-coupled and spatially indirect exciton states. Here, we demonstrate power-dependent Rabi oscillations of direct excitons, spatially indirect excitons, and excitons with a hybridized electron wave function. An off-resonant detection technique based on phonon-mediated state transfer allows for spectrally filtered detection under resonant excitation. Applying a gate voltage to the QDM-device enables a continuous transition between direct and indirect excitons and, thereby, control of the overlap of the electron and hole wave function. This does not only vary the Rabi frequency of the investigated transition by a factor of $\approx3$, but also allows to optimize graph state generation in terms of optical pulse power and reduction of radiative lifetimes.

\end{abstract}

\maketitle


\section{\label{sec:Intro}Introduction}
The use of single photons as flying qubits facilitates transmission of quantum information at the speed of light. However, transfer over large distances unavoidably comes with losses and decoherence. Encoding quantum information on an ensemble of entangled photons, a so-called graph state\,\cite{Briegel2001}, instead of a single photon, provides a possibility to mitigate the losses is transmission channels\,\cite{Azuma2015, Azuma2017}. Furthermore, other specific forms of graph states such as photonic cluster states promise realization of measurement-based quantum computing\,\cite{Raussendorf2001} as well as quantum error correction\,\cite{Schlingemann2002,Bell2014}. 

Following the Lindner-Rudolph protocol\,\cite{Lindner2009}, one-dimensional photonic cluster states can be deterministically generated by utilizing single spins in semiconductor quantum dots (QDs). The polarization entanglement of up to five photons has been achieved in a one-dimensional cluster state has been achieved\,\cite{Schwartz2016} and most recent experiments demonstrate localizable entanglement over ten photons\,\cite{Cogan2021}. While the nanophotonic environment of QDs provides high photon emission rates, the cluster state creation fidelity is limited by spin dephasing and modified selection rules in the presence of a transverse magnetic field\,\cite{Cogan2021}. These challenges can be overcome by using a pair of tunnel coupled and vertically stacked QDs, so called quantum dot molecules (QDMs)\,\cite{Vezvaee2022}. Besides prolonging the spin coherence compared to single quantum dots\,\cite{Tran2022}, QDMs possess an unique level structure\,\cite{Doty2010}. This level structure enables, for example, spin rotations and spin readout transitions without application of a magnetic field. The ability to create spatially indirect excitons, with one charge carrier occupying the upper and one the lower QD\,\cite{Krenner2005}, provides a cycling transition which can be used for generating time-bin entangled photons\,\cite{Vezvaee2022}. Moreover, QDMs are proposed to generate two-dimensional photonic cluster states by harnessing the tunnel coupling between the two QDs and inter-dot control gates\,\cite{Economou2010}.

The foundation for creating one- and two-dimensional photonic cluster states is the occurrence of excitons in spatially direct, spatially indirect, and hybridized configurations\,\cite{Stinaff2006}. In these different configurations, the charge carriers of an electron-hole pair are located in the same QD, in different QDs, or one of the charge wave functions is hybridized over both quantum dots, respectively. In each configuration, the overlap of the electron and hole wave functions and, therefore, the transition dipole moment (TDM) of the corresponding optical transition differs. This results in a change of both the lifetime of the excited state and the pulse area needed for maximal population inversion\,\cite{Wijesundara2011}. 
While the lifetime influences the cluster state creation efficiency and rate, the $\pi$-pulse area sets the intensity of the required optical control pulses. Hence, the TDM of the addressed transitions influences the generation process of photonic cluster states.
Furthermore, the proposed protocols require coherent excitation of electron-hole pairs in various exciton configurations to control and readout the exciton spin state.

In this work, we demonstrate coherent Rabi oscillations of direct, spatially indirect, and hybridized excitons in a single QDM. An off-resonant detection technique is introduced and applied, relying on phonon-mediated state transfers. We examine the dependence of the Rabi frequency on the excitonic configuration, as the overlap of the electron and hole wave functions changes. Tuning the electric field via a gate voltage allows electrical control of this wave function overlap and, therefore, of the pulse area needed for population inversion. In this way, we demonstrate and quantify electric control of the TDM. Finally, a simple one-dimensional model of a double-well potential allows us to model the voltage-dependence of the TDM. 


\begin{figure}
\includegraphics{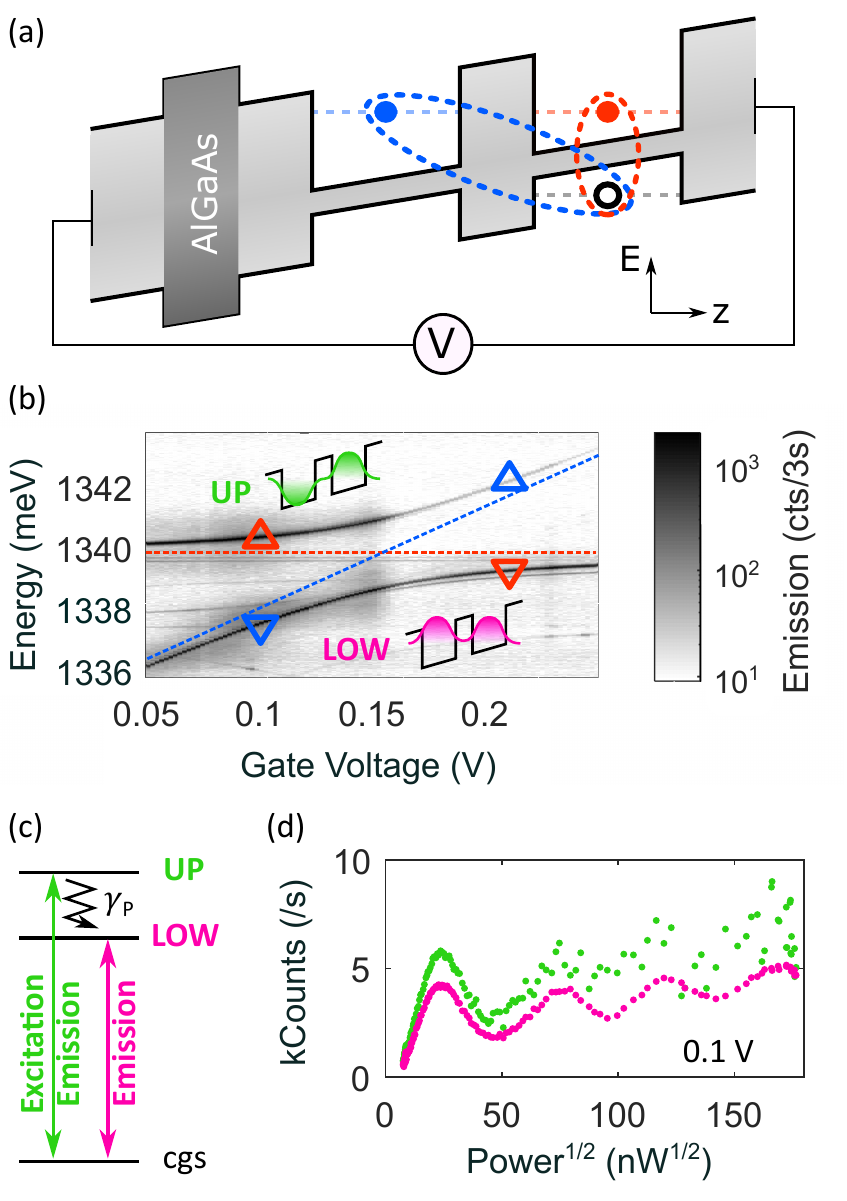}
\caption{\label{fig:1}Rabi oscillations of the neutral exciton in a QDM. (a)\,Schematic band structure of a QDM represented by a double-well potential. An AlGaAs barrier below the molecule prolongs tunneling times for electrons while not affecting tunneling for holes. One hole (empty circle) is located in the upper QD, while electrons (filled circles) occur in both dots. As a consequence, direct (red ellipse) and indirect (blue ellipse) excitons arise. A gate voltage V applied to the sample facilitates tuning of the direct and indirect exciton energies relative to each other. (b)\,Voltage-dependent photoluminescence of the neutral exciton. The red and blue dashed lines indicate the energies of the direct and indirect excitons. tunnel coupling between the two QDs leads to an avoided crossing with a symmetric (pink) and an anti-symmetric (green) electron eigenstate. The upper (lower) energy transition is called UP (LOW). Triangles indicate the excitation energy and voltage applied in Figure \ref{fig:2}. (c)\,Neutral exciton state diagram illustrating the excitation and detection scheme for monitoring Rabi oscillations. While a resonant light field (green) is driving UP, a phonon-mediated state transfer with rate $\gamma_P$ (black arrow) is enabling emission from both UP and the energetically detuned LOW. (d)\,Power-dependent Rabi oscillations when exciting UP and detecting UP (green) or LOW (pink) at 0.1\,V.}
\end{figure}

\section{\label{sec:level1}Results}

By vertically stacking two QDs with a separation in the nm regime, charge wave functions can hybridize across both QDss. In addition, both direct and spatially indirect excitons can form. Figure \ref{fig:1}\,(a) illustrates a schematic band-diagram of a QDM. The two QDs are depicted by a double-well potential, in which electrons (filled circle) and holes (empty circle) are trapped. The design of the investigated sample, described in Appendix \ref{sec:Sample}, energetically favours the location of a hole in the top QD. Consequently, a direct/indirect exciton (red/blue ellipse) forms, when an electron is trapped in the top/bottom QD. The QDM is embedded in a p-i-n diode structure; applying a gate voltage V facilitates tuning of the energy levels of both QDs relative to each other. In this way, the direct and indirect exciton energies can be brought into resonance. At the resonance condition, the electron wave function hybridizes across both dots, molecular bonding and anti-bonding states form, and an avoided crossing between the orbital states occurs. Since we can control the tunnel coupling between the two QDs by varying the gate voltage, we use this dependency to investigate coherent driving of different exciton configurations.

The most elemental charge state exhibiting the hybridization of wave functions is the neutral exciton ($X^0$). Figure \ref{fig:1}\,(b) shows a voltage-dependent photoluminescence measurement of the $X^0$. We make use of a two-phase electrical and optical sequence to deterministically prepare the QDM in a zero-charge ground-state and individually adjust the tunnel coupling\,\cite{Bopp2022}. Exciting the energetically higher p-shell orbital of the upper dot at 1353.6\,meV enables the unimpeded detection of the $X^0$ s-shell emission for multiple coupling conditions. At 0.16\,V, the electron wave function hybridizes and an avoided crossing forms. The resulting electron eigenstates are described by symmetric and antisymmetric wave functions\,\cite{Krenner2005}. The corresponding lower and higher energy transitions of the avoided crossing are denoted LOW and UP in Figure \ref{fig:1}\,(b). The red and blue dashed lines depict the energies of a direct and indirect exciton, respectively. By increasing the gate voltage, the exciton character changes from direct to hybridized to indirect for the upper energy branch, and vice versa for the lower energy branch. As a result, the overlap of the electron and hole wave functions changes.

The change of the wave function overlap is quantified by coherently driving Rabi oscillations on the exciton transition. The Rabi frequency of a resonantly excited two-level system $\Omega_R=\left|\frac{E_0D}{\hbar}\right|$ is linearly dependent on the TDM $D$, which in return is proportional to the overlap of the electron and hole wave function\,\cite{Fox1970}. In addition, $\Omega_R$ depends linearly on the electric driving field amplitude $E_0$. The $E_0$-dependence allows the observation of power-dependent Rabi oscillations\,\cite{Stievater2001}. For this purpose, a 5\,ps laser pulse is applied to resonantly drive the crystal ground state (cgs)-to-$X^0$ transition in the QDM. The occupation of the excited state is monitored by detecting the photons emitted by the driven two-level system. Commonly, emission from resonantly excited states is detected in a cross-polarized setup configuration to suppress the excitation laser\,\cite{Kuhlmann2013}. At high excitation power, however, laser light can leak into the detection channel and reduce the signal-to-noise ratio. We propose and demonstrate a readout technique utilizing a phonon-mediated state transfer\,\cite{Nakaoka2006}, which detunes the emitted photons energetically from the two-level system. Thereby, the limitation of an insufficiently suppressed excitation laser is eliminated via spectral filtering, and the visibility of the Rabi oscillations is increased.

Figure\,\ref{fig:1}\,(c) visualizes the state diagram of the $X^0$. The two excited states UP and LOW can both radiatively decay into the cgs. A phonon emission process with rate $\gamma_P$ can transfer the electron from the UP to the LOW configuration\,\cite{Nakaoka2006}. Since the excitation pulse length is short compared to the decay rates, the cgs-UP system is well approximated by a two-level system. It is coherently driven by a 5 ps laser pulse (green arrow). Figure\,\ref{fig:1}\,(d) shows the power-dependent resonance fluorescence emission of the UP transition as green data points. The measurement is performed at 0.1\,V, such that the driven transition exhibits a direct exciton character, as shown in Figure\,\ref{fig:1}\,(b). Rabi oscillations are observed up to a pulse area of slightly above $2\pi$ and $60^2$\,nW. However, a decreasing signal-to-noise ratio prevents the detection of oscillations above $60^2$\,nW due to nsufficient suppression of the excitation laser. 

To improve the signal-to-noise ratio, which decreases with increasing power, we make use of a phonon-mediated state transfer. The emission of a phonon transfers the electron from the UP into the LOW configuration. This process can only occur as long as the system is in the excited state. Thus, the ensemble occupation of LOW is proportional to the ensemble occupation of UP, and so is the number of emitted photons of both transitions. In addition, due to the avoided crossing, the emission of LOW is at least 2.1\,meV detuned from the driving energy for any gate voltage, which allows the spectral filtering of the emission from the excitation laser pulse. Thus, the resonant excitation of the two-level system and the off-resonant monitoring of its excited state occupation are achieved simultaneously. The power-dependent emission of the LOW transition when exciting UP is shown by the pink data points in Figure \ref{fig:1}\,(d). Below $60^2$\,nW, both readout techniques show the same Rabi frequency as expected, confirming the proportionality of occupancy between UP and LOW. However, in contrast to the resonant detection (green), Rabi oscillations are well resolvable up to a pulse area of $7\pi$. The reduction of the oscillation amplitude arises from interactions with phonons\,\cite{Forstner2003}, while the increase of the mean is attributed to a slightly chirped excitation laser pulse\,\cite{Kaldewey2017}. From the relative intensities of both transitions, we can conclude that the phonon induced relaxation rate is comparable to the radiative decay rate of the direct UP transition.\\


\begin{figure}
\includegraphics{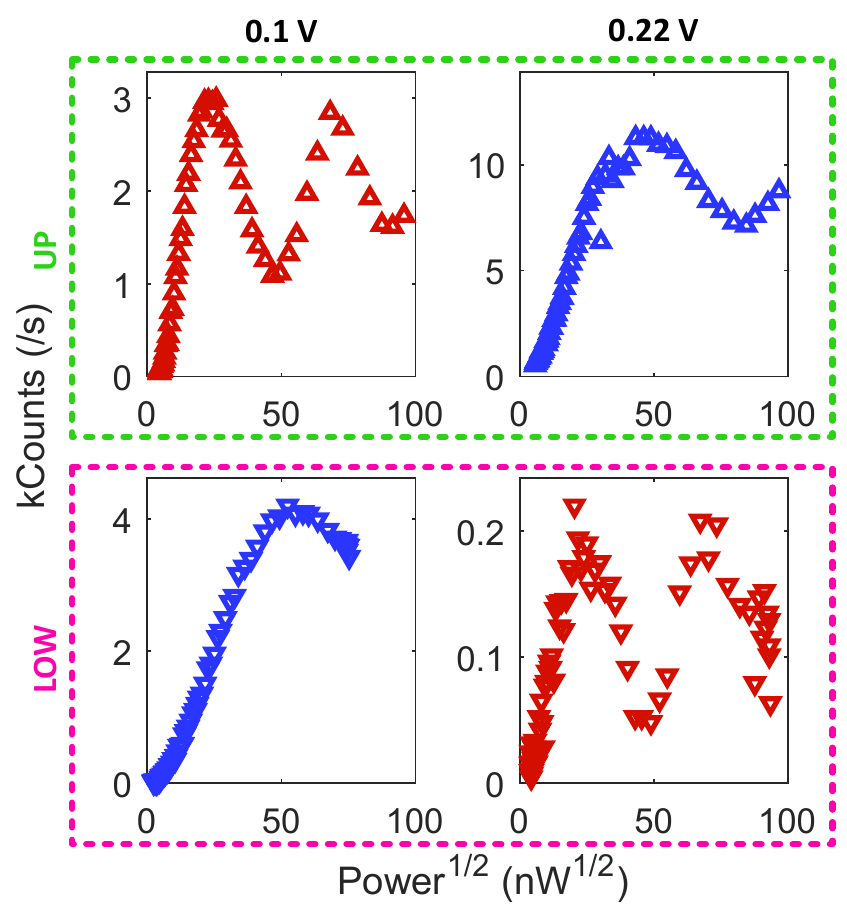}
\caption{\label{fig:2} Rabi oscillations of the UP and LOW branch at 0.1\,V (left) and 0.22 V (right) by phonon-mediated state transfers. The red data points correspond to a direct, the blue to an
indirect driven transition.}
\end{figure}

Electric control of the tunnel coupling between the two QDs allows coherent excitation of electron-hole pairs in different occupation configurations. Figure \ref{fig:2} shows the power-dependent emission of the QDM while resonantly exciting UP and detecting LOW (green dashed box, UP). The measurements are performed at 0.1\,V (left) and 0.22\,V (right), on either side of the avoided crossing. The red and blue data points indicate a direct and indirect character of the excited transition, respectively. We observe Rabi oscillations for both the direct and indirect transitions, which confirms that coherent excitation of a spatially indirect exciton is possible. However, the Rabi frequency of the indirect configuration is reduced compared to the direct configuration. This is caused by the reduced overlap of the electron and hole wave functions and the accompanying decrease of the TDM for the indirect exciton.

A verification of these results is found by performing the same experiments on the lower branch (Figure \ref{fig:2}, pink box, LOW). Similar to the previous case, a phonon absorption process facilitates a state transfer from LOW to UP and, therefore, the off-resonant detection of UP while exciting LOW. This allows us to off-resonantly monitor Rabi oscillations of the LOW branch. The investigated gate voltages of both excitation cases are chosen to be $\pm$0.06\,V away from the avoided crossing. Therefore, the electron configuration of the upper branch at 0.1\,V (0.22\,V) resembles the electron configuration of the lower branch at 0.22\,V (0.1\,V). This leads to a comparable Rabi frequency of the direct (red) and indirect (blue) exciton of the upper and lower energy transition at the two voltages. The difference in absolute counts between the excitation of UP and LOW is attributed to the underlying phonon process. When exciting UP (LOW), we rely on the emission (absorption) of a phonon to detect the signal. Since the measurements are performed at 10\,K, the probability of absorbing a phonon is strongly reduced compared to the emission.\\

\begin{figure}
\includegraphics{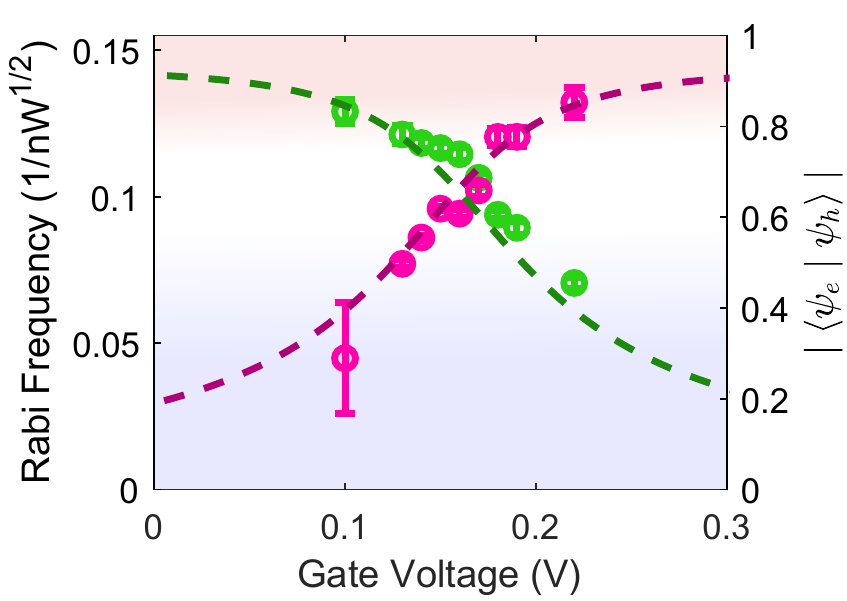}
\caption{\label{fig:3} Measured voltage dependent Rabi frequency of UP (green) and LOW (pink), plotted on the left axes. The right axes visualizes the calculated overlap of the electron and hole wave functions as a function of the voltage, where the pink/green dashed line corresponds to the lowest/second-lowest electron eigenenergy. The red/blue shaded background indicates the direct/indirect character of the transition.}
\end{figure}

Since our QDM device allows continuous tuning the gate voltage while maintaining the prepared charge state, arbitrary exciton configurations can be set. Thereby, the overlap of the electron and hole wave functions is analyzable for any coupling condition. Figure \ref{fig:3} shows the power-dependent Rabi frequencies as a function of the gate voltage for the upper (green) and lower branch (pink). The Rabi frequencies are extracted by fitting a $sin^2(\text{laser power})$ function to the data, with an exponential decay to take phonon dephasing into account, and a linear increase with intensity to compensate for a chirped excitation pulse. We observe a continuous increase (decrease) of the frequency when transitioning from an indirect (direct) to a direct (indirect) exciton. By raising the gate voltage and following the UP transition, the electron occupation shifts from the top to the bottom dot. The opposite holds for the LOW transition. This leads to a continuous variation of the overlap of the electron and hole wave functions and, consequently, to a change in the Rabi frequency. Within the investigated range between 0.1\,V and 0.22\,V, we are able to electrically tune the Rabi frequency by a factor of $\approx3$.

The wave function overlap is modeled by calculating the eigenenergies and -values of a tilted, one-dimensional double-squarewell potential representing the QDM. By fitting the energy difference between the two lowest eigenstates to the voltage-dependent separation of UP and LOW, the depth of the squarewell potential and the effective electron mass are determined. A detailed description of the model is provided in Appendix \ref{sec:Model}. The right axes of Figure \ref{fig:3} shows the overlap of the electron and hole wave functions $|\langle\psi_e|\psi_h\rangle|$ for the lowest (pink) and second lowest (green) electron eigenenergy by dashed lines. The electron eigenenergies correspond to the LOW and UP transition, respectively. The one-dimensional model provides a remarkably good description of the measured voltage-dependent Rabi frequencies. Thereby, the Rabi frequency can be related to the TDM of direct, indirect, and hybridized excitons, which allows determination the $\pi$-pulse area as well as the difference in radiative lifetime of the corresponding transition.

\section{\label{sec:discussion}Discussion and summary}

Adressing direct, indirect, and hybridized excitons is fundamental for using QDMs as spin-photon interfaces. In addition, the electrical tuneability of the TDM of the adressed transitions at and around the tunnel coupling regime is one key parameter of a QDM. It not only determines the $\pi$-pulse power of the addressed transition, as shown in this work, but is also directly related to the lifetime of the excited state. Therefore, it is one of the parameters setting the creation rate for generating one- and two- dimensional photonic cluster states as well as for performing quantum-repeater protocols. 

We have demonstrated the coherent excitation of direct, indirect, and hybridized excitons -- one of the elementary building blocks for creating photonic cluster states from QDMs. We use non-resonant readout, which is facilitated by phonon-mediated charge relaxation and excitation between the two lowest energy eigenstates of the electron.. Voltage-dependent Rabi oscillations show a continuous increase of the Rabi frequency when transitioning from an indirect to a direct exciton. This is attributed to an electrically controlled increase of the TDM of a direct compared to an indirect transition. Furthermore, we apply a one-dimensional model to calculate the overlap of the $X^0$ electron and hole wave functions. Within the voltage range presented, we are able to tune the Rabi frequency and consequently the TDM by a factor of $\approx3$. This corresponds to a variation of the radiative lifetime between a direct and an indirect exciton by a factor of $\approx9$, as it scales quadratically with the TDM.

The coherent excitation and the electrical tunability between various exciton configurations in QDMs not only paves the way towards the generation of entangled multi-photon states. It might also enable protocols which utilize fast electrical switching between the exciton configurations. This can reduce their lifetime and the $\pi$-pulse power and highly improve the cluster state generation rate.

\begin{acknowledgments}
The authors gratefully acknowledge financial support from the German Federal Ministry of Education and Research (BMBF) via Q.Link.X (16KIS0874, 16KIS086), QR.X (16KISQ027, 16KISQ014, 16KISQ012 and 16KISQ009), the European Union’s Horizon 2020 research and innovation program under grant agreement 862035 (QLUSTER) and the Deutsche Forschungsgemeinschaft (DFG, German Research Foundation) via SQAM (FI947-5-1), DIP (FI947-6-1), and the Excellence Cluster MCQST (EXC-2111, 390814868). F.B. gratefully acknowledges the Exploring Quantum Matter (ExQM) programme funded by the State of Bavaria. F.S., K.B., and K.M. gratefully acknowledge the BMBF for financial support via project MOQUA (13N14846).
\end{acknowledgments}

\appendix
\section{Sample}
\label{sec:Sample}

The investigated InAs QDM is enclosed in a GaAs matrix and was grown by molecular beam epitaxy. It consists of two vertically stacked QDs. The inter-dot coupling strength is determined by a wetting layer-to-wetting layer separation of 10\,nm. In addition, an Al$_x$Ga$_{(x-1)}$As barrier ($x=0.33$) with a thickness of 2.5\,nm is placed between the dots to reduce the coupling strength. The height of the top (bottom) QD was fixed to 2.9\,nm (2.7\,nm) via the indium-flush technique during growth. This height configuration facilitates electric field-induced tunnel coupling of orbital states in the conduction band. A 50\,nm thick Al$_x$Ga$_{(x-1)}$As tunnel barrier ($x=0.33$) was grown 5 nm below the QDM to prolong electron tunneling times. The molecule is embedded in a p-i-n diode, with the doped regions used as contacts to gate the sample. The diode contacts are placed more than 150\,nm away from the molecule to prevent uncontrolled charge tunneling into the QDM. Furthermore, a distributed Bragg reflector was grown below the diode and a circular Bragg grating was positioned deterministically  via in-situ electron beam lithography above an individual and pre-selected QDM to improve photon in- and outcoupling efficiencies\,\cite{Schall2021}. All measurements are performed at 10 K.\\
\section{Double-well potential model}
\label{sec:Model}

To calculate the overlap of the electron and hole wave functions, we set up a model consisting of a double-squarewell potential representing the conduction band of the QDM. We assume that the variation of the in-plane wave functions is small compared to the wave functions along the growth direction $z$, when changing the gate voltage. This assumption is reasonable, since the confinement of charges along the growth axes and the translation introduced by the gate voltage along the growth axes exceeds the in-plane variation. We can, therefore, approach the problem with a one-dimensional model and expect an acceptable degree of accuracy. The potential $V(z)$ is designed to match the dimensions of the QDM with respect to the tunnel barrier width and dot heights (see Section \ref{sec:Sample}). $z$ is here the growth direction of the sample. In addition, we tilt the potential to imitate the presence of an applied gate voltage. Solving the time-independent Schrödinger equation for a given gate voltage allows us to obtain the envelope functions $\Psi$ and their eigenenergies $E$ of an electron with mass $m_e$ trapped in the double-well potential. The envelope functions are then used to represent the wave function of the electron since the Bloch part of the wave functions are only weakly sensitive to electric fields of the magnitude applied here.

To define the free parameters of the double-well model, we determine the effective electron mass $m_e$ and the potential depth by fitting the difference between the calculated two lowest eigenenergies to the measured energy difference $\Delta$E between UP and LOW. $\Delta$E between the two $X^0$ branches is purely determined by the energy difference between the electron eigenstates. For calculating the hole wave function, the potential is inverted to represent the valance band. Its depth is set to match half the depth of the electron potential whereas the heavy hole mass is set to match $m_h=10\,m_e$\,\cite{Bouarissa1999}. Since we are interested in calculating the overlap of wave functions rather than absolute transition energies, it is sufficient to work with relative values in the model.

\bibliography{apssamp}

\providecommand{\noopsort}[1]{}\providecommand{\singleletter}[1]{#1}%
\begin{thebibliography}{25}%
\makeatletter
\providecommand \@ifxundefined [1]{%
 \@ifx{#1\undefined}
}%
\providecommand \@ifnum [1]{%
 \ifnum #1\expandafter \@firstoftwo
 \else \expandafter \@secondoftwo
 \fi
}%
\providecommand \@ifx [1]{%
 \ifx #1\expandafter \@firstoftwo
 \else \expandafter \@secondoftwo
 \fi
}%
\providecommand \natexlab [1]{#1}%
\providecommand \enquote  [1]{``#1''}%
\providecommand \bibnamefont  [1]{#1}%
\providecommand \bibfnamefont [1]{#1}%
\providecommand \citenamefont [1]{#1}%
\providecommand \href@noop [0]{\@secondoftwo}%
\providecommand \href [0]{\begingroup \@sanitize@url \@href}%
\providecommand \@href[1]{\@@startlink{#1}\@@href}%
\providecommand \@@href[1]{\endgroup#1\@@endlink}%
\providecommand \@sanitize@url [0]{\catcode `\\12\catcode `\$12\catcode
  `\&12\catcode `\#12\catcode `\^12\catcode `\_12\catcode `\%12\relax}%
\providecommand \@@startlink[1]{}%
\providecommand \@@endlink[0]{}%
\providecommand \url  [0]{\begingroup\@sanitize@url \@url }%
\providecommand \@url [1]{\endgroup\@href {#1}{\urlprefix }}%
\providecommand \urlprefix  [0]{URL }%
\providecommand \Eprint [0]{\href }%
\providecommand \doibase [0]{https://doi.org/}%
\providecommand \selectlanguage [0]{\@gobble}%
\providecommand \bibinfo  [0]{\@secondoftwo}%
\providecommand \bibfield  [0]{\@secondoftwo}%
\providecommand \translation [1]{[#1]}%
\providecommand \BibitemOpen [0]{}%
\providecommand \bibitemStop [0]{}%
\providecommand \bibitemNoStop [0]{.\EOS\space}%
\providecommand \EOS [0]{\spacefactor3000\relax}%
\providecommand \BibitemShut  [1]{\csname bibitem#1\endcsname}%
\let\auto@bib@innerbib\@empty
\bibitem [{\citenamefont {Briegel}\ and\ \citenamefont
  {Raussendorf}(2001)}]{Briegel2001}%
  \BibitemOpen
  \bibfield  {author} {\bibinfo {author} {\bibfnamefont {H.~J.}\ \bibnamefont
  {Briegel}}\ and\ \bibinfo {author} {\bibfnamefont {R.}~\bibnamefont
  {Raussendorf}},\ }\bibfield  {title} {\bibinfo {title} {{Persistent
  entanglement in arrays of interacting particles}},\ }\href
  {https://journals.aps.org/prl/abstract/10.1103/PhysRevLett.86.910} {\bibfield
   {journal} {\bibinfo  {journal} {Physical Review Letters}\ }\textbf {\bibinfo
  {volume} {5}},\ \bibinfo {pages} {910} (\bibinfo {year} {2001})}\BibitemShut
  {NoStop}%
\bibitem [{\citenamefont {Azuma}\ \emph {et~al.}(2015)\citenamefont {Azuma},
  \citenamefont {Tamaki},\ and\ \citenamefont {Lo}}]{Azuma2015}%
  \BibitemOpen
  \bibfield  {author} {\bibinfo {author} {\bibfnamefont {K.}~\bibnamefont
  {Azuma}}, \bibinfo {author} {\bibfnamefont {K.}~\bibnamefont {Tamaki}},\ and\
  \bibinfo {author} {\bibfnamefont {H.~K.}\ \bibnamefont {Lo}},\ }\bibfield
  {title} {\bibinfo {title} {{All-photonic quantum repeaters}},\ }\href@noop {}
  {\bibfield  {journal} {\bibinfo  {journal} {Nature Communications}\ }\textbf
  {\bibinfo {volume} {6}} (\bibinfo {year} {2015})}\BibitemShut {NoStop}%
\bibitem [{\citenamefont {Azuma}\ and\ \citenamefont {Kato}(2017)}]{Azuma2017}%
  \BibitemOpen
  \bibfield  {author} {\bibinfo {author} {\bibfnamefont {K.}~\bibnamefont
  {Azuma}}\ and\ \bibinfo {author} {\bibfnamefont {G.}~\bibnamefont {Kato}},\
  }\bibfield  {title} {\bibinfo {title} {{Aggregating quantum repeaters for the
  quantum internet}},\ }\href
  {https://journals.aps.org/pra/abstract/10.1103/PhysRevA.96.032332} {\bibfield
   {journal} {\bibinfo  {journal} {Physical Review A}\ }\textbf {\bibinfo
  {volume} {96}},\ \bibinfo {pages} {032332} (\bibinfo {year}
  {2017})}\BibitemShut {NoStop}%
\bibitem [{\citenamefont {Raussendorf}\ and\ \citenamefont
  {Briegel}(2001)}]{Raussendorf2001}%
  \BibitemOpen
  \bibfield  {author} {\bibinfo {author} {\bibfnamefont {R.}~\bibnamefont
  {Raussendorf}}\ and\ \bibinfo {author} {\bibfnamefont {H.~J.}\ \bibnamefont
  {Briegel}},\ }\bibfield  {title} {\bibinfo {title} {{A one-way quantum
  computer}},\ }\href@noop {} {\bibfield  {journal} {\bibinfo  {journal}
  {Physical Review Letters}\ }\textbf {\bibinfo {volume} {86}},\ \bibinfo
  {pages} {5188} (\bibinfo {year} {2001})}\BibitemShut {NoStop}%
\bibitem [{\citenamefont {Schlingemann}\ and\ \citenamefont
  {Werner}(2002)}]{Schlingemann2002}%
  \BibitemOpen
  \bibfield  {author} {\bibinfo {author} {\bibfnamefont {D.}~\bibnamefont
  {Schlingemann}}\ and\ \bibinfo {author} {\bibfnamefont {R.~F.}\ \bibnamefont
  {Werner}},\ }\bibfield  {title} {\bibinfo {title} {{Quantum error-correcting
  codes associated with graphs}},\ }\href
  {https://journals.aps.org/pra/abstract/10.1103/PhysRevA.65.012308} {\bibfield
   {journal} {\bibinfo  {journal} {Physical Review A - Atomic, Molecular, and
  Optical Physics}\ }\textbf {\bibinfo {volume} {65}},\ \bibinfo {pages} {8}
  (\bibinfo {year} {2002})}\BibitemShut {NoStop}%
\bibitem [{\citenamefont {Bell}\ \emph {et~al.}(2014)\citenamefont {Bell},
  \citenamefont {Herrera-Mart{\'{i}}}, \citenamefont {Tame}, \citenamefont
  {Markham}, \citenamefont {Wadsworth},\ and\ \citenamefont
  {Rarity}}]{Bell2014}%
  \BibitemOpen
  \bibfield  {author} {\bibinfo {author} {\bibfnamefont {B.~A.}\ \bibnamefont
  {Bell}}, \bibinfo {author} {\bibfnamefont {D.~A.}\ \bibnamefont
  {Herrera-Mart{\'{i}}}}, \bibinfo {author} {\bibfnamefont {M.~S.}\
  \bibnamefont {Tame}}, \bibinfo {author} {\bibfnamefont {D.}~\bibnamefont
  {Markham}}, \bibinfo {author} {\bibfnamefont {W.~J.}\ \bibnamefont
  {Wadsworth}},\ and\ \bibinfo {author} {\bibfnamefont {J.~G.}\ \bibnamefont
  {Rarity}},\ }\bibfield  {title} {\bibinfo {title} {{Experimental
  demonstration of a graph state quantum error-correction code}},\ }\href
  {www.nature.com/naturecommunications} {\bibfield  {journal} {\bibinfo
  {journal} {Nature Communications}\ }\textbf {\bibinfo {volume} {5}},\
  \bibinfo {pages} {1} (\bibinfo {year} {2014})}\BibitemShut {NoStop}%
\bibitem [{\citenamefont {Lindner}\ and\ \citenamefont
  {Rudolph}(2009)}]{Lindner2009}%
  \BibitemOpen
  \bibfield  {author} {\bibinfo {author} {\bibfnamefont {N.~H.}\ \bibnamefont
  {Lindner}}\ and\ \bibinfo {author} {\bibfnamefont {T.}~\bibnamefont
  {Rudolph}},\ }\bibfield  {title} {\bibinfo {title} {{Proposal for pulsed
  On-demand sources of photonic cluster state strings}},\ }\href@noop {}
  {\bibfield  {journal} {\bibinfo  {journal} {Physical Review Letters}\
  }\textbf {\bibinfo {volume} {103}} (\bibinfo {year} {2009})}\BibitemShut
  {NoStop}%
\bibitem [{\citenamefont {Schwartz}\ \emph {et~al.}(2016)\citenamefont
  {Schwartz}, \citenamefont {Cogan}, \citenamefont {Schmidgall}, \citenamefont
  {Don}, \citenamefont {Gantz}, \citenamefont {Kenneth}, \citenamefont
  {Lindner},\ and\ \citenamefont {Gershoni}}]{Schwartz2016}%
  \BibitemOpen
  \bibfield  {author} {\bibinfo {author} {\bibfnamefont {I.}~\bibnamefont
  {Schwartz}}, \bibinfo {author} {\bibfnamefont {D.}~\bibnamefont {Cogan}},
  \bibinfo {author} {\bibfnamefont {E.~R.}\ \bibnamefont {Schmidgall}},
  \bibinfo {author} {\bibfnamefont {Y.}~\bibnamefont {Don}}, \bibinfo {author}
  {\bibfnamefont {L.}~\bibnamefont {Gantz}}, \bibinfo {author} {\bibfnamefont
  {O.}~\bibnamefont {Kenneth}}, \bibinfo {author} {\bibfnamefont {N.~H.}\
  \bibnamefont {Lindner}},\ and\ \bibinfo {author} {\bibfnamefont
  {D.}~\bibnamefont {Gershoni}},\ }\bibfield  {title} {\bibinfo {title}
  {{Deterministic generation of a cluster state of entangled photons}},\ }\href
  {https://www.science.org} {\bibfield  {journal} {\bibinfo  {journal}
  {Science}\ }\textbf {\bibinfo {volume} {354}},\ \bibinfo {pages} {434}
  (\bibinfo {year} {2016})}\BibitemShut {NoStop}%
\bibitem [{\citenamefont {Cogan}\ \emph {et~al.}(2021)\citenamefont {Cogan},
  \citenamefont {Su}, \citenamefont {Kenneth},\ and\ \citenamefont
  {Gershoni}}]{Cogan2021}%
  \BibitemOpen
  \bibfield  {author} {\bibinfo {author} {\bibfnamefont {D.}~\bibnamefont
  {Cogan}}, \bibinfo {author} {\bibfnamefont {Z.-E.}\ \bibnamefont {Su}},
  \bibinfo {author} {\bibfnamefont {O.}~\bibnamefont {Kenneth}},\ and\ \bibinfo
  {author} {\bibfnamefont {D.}~\bibnamefont {Gershoni}},\ }\bibfield  {title}
  {\bibinfo {title} {{A deterministic source of indistinguishable photons in a
  cluster state}},\ }\href {http://arxiv.org/abs/2110.05908} {\  (\bibinfo
  {year} {2021})},\ \Eprint {https://arxiv.org/abs/2110.05908}
  {arXiv:2110.05908} \BibitemShut {NoStop}%
\bibitem [{\citenamefont {Vezvaee}\ \emph {et~al.}(2022)\citenamefont
  {Vezvaee}, \citenamefont {Hilaire}, \citenamefont {Doty},\ and\ \citenamefont
  {Economou}}]{Vezvaee2022}%
  \BibitemOpen
  \bibfield  {author} {\bibinfo {author} {\bibfnamefont {A.}~\bibnamefont
  {Vezvaee}}, \bibinfo {author} {\bibfnamefont {P.}~\bibnamefont {Hilaire}},
  \bibinfo {author} {\bibfnamefont {M.~F.}\ \bibnamefont {Doty}},\ and\
  \bibinfo {author} {\bibfnamefont {S.~E.}\ \bibnamefont {Economou}},\
  }\bibfield  {title} {\bibinfo {title} {{Deterministic generation of entangled
  photonic cluster states from quantum dot molecules}},\ }\href
  {http://arxiv.org/abs/2206.03647} {\  (\bibinfo {year} {2022})},\ \Eprint
  {https://arxiv.org/abs/2206.03647} {arXiv:2206.03647} \BibitemShut {NoStop}%
\bibitem [{\citenamefont {Tran}\ \emph {et~al.}(2022)\citenamefont {Tran},
  \citenamefont {Bracker}, \citenamefont {Yakes}, \citenamefont {Grim},\ and\
  \citenamefont {Carter}}]{Tran2022}%
  \BibitemOpen
  \bibfield  {author} {\bibinfo {author} {\bibfnamefont {K.~X.}\ \bibnamefont
  {Tran}}, \bibinfo {author} {\bibfnamefont {A.~S.}\ \bibnamefont {Bracker}},
  \bibinfo {author} {\bibfnamefont {M.~K.}\ \bibnamefont {Yakes}}, \bibinfo
  {author} {\bibfnamefont {J.~Q.}\ \bibnamefont {Grim}},\ and\ \bibinfo
  {author} {\bibfnamefont {S.~G.}\ \bibnamefont {Carter}},\ }\bibfield  {title}
  {\bibinfo {title} {{Enhanced Spin Coherence of a Self-Assembled Quantum Dot
  Molecule at the Optimal Electrical Bias}},\ }\href
  {https://link.aps.org/doi/10.1103/PhysRevLett.129.027403} {\bibfield
  {journal} {\bibinfo  {journal} {Physical Review Letters}\ }\textbf {\bibinfo
  {volume} {129}},\ \bibinfo {pages} {2202.08256} (\bibinfo {year}
  {2022})}\BibitemShut {NoStop}%
\bibitem [{\citenamefont {Doty}\ \emph {et~al.}(2010)\citenamefont {Doty},
  \citenamefont {Climente}, \citenamefont {Greilich}, \citenamefont {Yakes},
  \citenamefont {Bracker},\ and\ \citenamefont {Gammon}}]{Doty2010}%
  \BibitemOpen
  \bibfield  {author} {\bibinfo {author} {\bibfnamefont {M.~F.}\ \bibnamefont
  {Doty}}, \bibinfo {author} {\bibfnamefont {J.~I.}\ \bibnamefont {Climente}},
  \bibinfo {author} {\bibfnamefont {A.}~\bibnamefont {Greilich}}, \bibinfo
  {author} {\bibfnamefont {M.}~\bibnamefont {Yakes}}, \bibinfo {author}
  {\bibfnamefont {A.~S.}\ \bibnamefont {Bracker}},\ and\ \bibinfo {author}
  {\bibfnamefont {D.}~\bibnamefont {Gammon}},\ }\bibfield  {title} {\bibinfo
  {title} {{Hole-spin mixing in InAs quantum dot molecules}},\ }\href@noop {}
  {\bibfield  {journal} {\bibinfo  {journal} {Physical Review B - Condensed
  Matter and Materials Physics}\ }\textbf {\bibinfo {volume} {81}} (\bibinfo
  {year} {2010})}\BibitemShut {NoStop}%
\bibitem [{\citenamefont {Krenner}\ \emph {et~al.}(2005)\citenamefont
  {Krenner}, \citenamefont {Sabathil}, \citenamefont {Clark}, \citenamefont
  {Kress}, \citenamefont {Schuh}, \citenamefont {Bichler}, \citenamefont
  {Abstreiter},\ and\ \citenamefont {Finley}}]{Krenner2005}%
  \BibitemOpen
  \bibfield  {author} {\bibinfo {author} {\bibfnamefont {H.~J.}\ \bibnamefont
  {Krenner}}, \bibinfo {author} {\bibfnamefont {M.}~\bibnamefont {Sabathil}},
  \bibinfo {author} {\bibfnamefont {E.~C.}\ \bibnamefont {Clark}}, \bibinfo
  {author} {\bibfnamefont {A.}~\bibnamefont {Kress}}, \bibinfo {author}
  {\bibfnamefont {D.}~\bibnamefont {Schuh}}, \bibinfo {author} {\bibfnamefont
  {M.}~\bibnamefont {Bichler}}, \bibinfo {author} {\bibfnamefont
  {G.}~\bibnamefont {Abstreiter}},\ and\ \bibinfo {author} {\bibfnamefont
  {J.~J.}\ \bibnamefont {Finley}},\ }\bibfield  {title} {\bibinfo {title}
  {{Direct Observation of Controlled Coupling in an Individual Quantum Dot
  Molecule}},\ }\href {https://link.aps.org/doi/10.1103/PhysRevLett.94.057402}
  {\bibfield  {journal} {\bibinfo  {journal} {Physical Review Letters}\
  }\textbf {\bibinfo {volume} {94}},\ \bibinfo {pages} {057402} (\bibinfo
  {year} {2005})}\BibitemShut {NoStop}%
\bibitem [{\citenamefont {Economou}\ \emph {et~al.}(2010)\citenamefont
  {Economou}, \citenamefont {Lindner},\ and\ \citenamefont
  {Rudolph}}]{Economou2010}%
  \BibitemOpen
  \bibfield  {author} {\bibinfo {author} {\bibfnamefont {S.~E.}\ \bibnamefont
  {Economou}}, \bibinfo {author} {\bibfnamefont {N.}~\bibnamefont {Lindner}},\
  and\ \bibinfo {author} {\bibfnamefont {T.}~\bibnamefont {Rudolph}},\
  }\bibfield  {title} {\bibinfo {title} {{Optically generated 2-dimensional
  photonic cluster state from coupled quantum dots}},\ }\href@noop {}
  {\bibfield  {journal} {\bibinfo  {journal} {Physical Review Letters}\
  }\textbf {\bibinfo {volume} {105}} (\bibinfo {year} {2010})}\BibitemShut
  {NoStop}%
\bibitem [{\citenamefont {Stinaff}\ \emph {et~al.}(2006)\citenamefont
  {Stinaff}, \citenamefont {Scheibner}, \citenamefont {Bracker}, \citenamefont
  {Ponomarev}, \citenamefont {Korenev}, \citenamefont {Ware}, \citenamefont
  {Doty}, \citenamefont {Reinecke},\ and\ \citenamefont
  {Gammon}}]{Stinaff2006}%
  \BibitemOpen
  \bibfield  {author} {\bibinfo {author} {\bibfnamefont {E.~A.}\ \bibnamefont
  {Stinaff}}, \bibinfo {author} {\bibfnamefont {M.}~\bibnamefont {Scheibner}},
  \bibinfo {author} {\bibfnamefont {A.~S.}\ \bibnamefont {Bracker}}, \bibinfo
  {author} {\bibfnamefont {I.~V.}\ \bibnamefont {Ponomarev}}, \bibinfo {author}
  {\bibfnamefont {V.~L.}\ \bibnamefont {Korenev}}, \bibinfo {author}
  {\bibfnamefont {M.~E.}\ \bibnamefont {Ware}}, \bibinfo {author}
  {\bibfnamefont {M.~F.}\ \bibnamefont {Doty}}, \bibinfo {author}
  {\bibfnamefont {T.~L.}\ \bibnamefont {Reinecke}},\ and\ \bibinfo {author}
  {\bibfnamefont {D.}~\bibnamefont {Gammon}},\ }\bibfield  {title} {\bibinfo
  {title} {{Optical Signatures of Coupled Quantum Dots}},\ }\href
  {http://www.sciencemag.org/cgi/doi/10.1126/science.1083800} {\bibfield
  {journal} {\bibinfo  {journal} {Science}\ }\textbf {\bibinfo {volume}
  {311}},\ \bibinfo {pages} {636} (\bibinfo {year} {2006})}\BibitemShut
  {NoStop}%
\bibitem [{\citenamefont {Wijesundara}\ \emph {et~al.}(2011)\citenamefont
  {Wijesundara}, \citenamefont {Rolon}, \citenamefont {Ulloa}, \citenamefont
  {Bracker}, \citenamefont {Gammon},\ and\ \citenamefont
  {Stinaff}}]{Wijesundara2011}%
  \BibitemOpen
  \bibfield  {author} {\bibinfo {author} {\bibfnamefont {K.~C.}\ \bibnamefont
  {Wijesundara}}, \bibinfo {author} {\bibfnamefont {J.~E.}\ \bibnamefont
  {Rolon}}, \bibinfo {author} {\bibfnamefont {S.~E.}\ \bibnamefont {Ulloa}},
  \bibinfo {author} {\bibfnamefont {A.~S.}\ \bibnamefont {Bracker}}, \bibinfo
  {author} {\bibfnamefont {D.}~\bibnamefont {Gammon}},\ and\ \bibinfo {author}
  {\bibfnamefont {E.~A.}\ \bibnamefont {Stinaff}},\ }\bibfield  {title}
  {\bibinfo {title} {{Tunable exciton relaxation in vertically coupled
  semiconductor InAs quantum dots}},\ }\href@noop {} {\bibfield  {journal}
  {\bibinfo  {journal} {Physical Review B - Condensed Matter and Materials
  Physics}\ }\textbf {\bibinfo {volume} {84}},\ \bibinfo {pages} {81404}
  (\bibinfo {year} {2011})}\BibitemShut {NoStop}%
\bibitem [{\citenamefont {Bopp}\ \emph {et~al.}(2022)\citenamefont {Bopp},
  \citenamefont {Rojas}, \citenamefont {Revenga}, \citenamefont {Riedl},
  \citenamefont {Sbresny}, \citenamefont {Boos}, \citenamefont {Simmet},
  \citenamefont {Ahmadi}, \citenamefont {Gershoni}, \citenamefont {Kasprzak},
  \citenamefont {Ludwig}, \citenamefont {Reitzenstein}, \citenamefont {Wieck},
  \citenamefont {Reuter}, \citenamefont {M{\"{u}}ller},\ and\ \citenamefont
  {Finley}}]{Bopp2022}%
  \BibitemOpen
  \bibfield  {author} {\bibinfo {author} {\bibfnamefont {F.}~\bibnamefont
  {Bopp}}, \bibinfo {author} {\bibfnamefont {J.}~\bibnamefont {Rojas}},
  \bibinfo {author} {\bibfnamefont {N.}~\bibnamefont {Revenga}}, \bibinfo
  {author} {\bibfnamefont {H.}~\bibnamefont {Riedl}}, \bibinfo {author}
  {\bibfnamefont {F.}~\bibnamefont {Sbresny}}, \bibinfo {author} {\bibfnamefont
  {K.}~\bibnamefont {Boos}}, \bibinfo {author} {\bibfnamefont {T.}~\bibnamefont
  {Simmet}}, \bibinfo {author} {\bibfnamefont {A.}~\bibnamefont {Ahmadi}},
  \bibinfo {author} {\bibfnamefont {D.}~\bibnamefont {Gershoni}}, \bibinfo
  {author} {\bibfnamefont {J.}~\bibnamefont {Kasprzak}}, \bibinfo {author}
  {\bibfnamefont {A.}~\bibnamefont {Ludwig}}, \bibinfo {author} {\bibfnamefont
  {S.}~\bibnamefont {Reitzenstein}}, \bibinfo {author} {\bibfnamefont
  {A.}~\bibnamefont {Wieck}}, \bibinfo {author} {\bibfnamefont
  {D.}~\bibnamefont {Reuter}}, \bibinfo {author} {\bibfnamefont
  {K.}~\bibnamefont {M{\"{u}}ller}},\ and\ \bibinfo {author} {\bibfnamefont
  {J.~J.}\ \bibnamefont {Finley}},\ }\bibfield  {title} {\bibinfo {title}
  {{Quantum Dot Molecule Devices with Optical Control of Charge Status and
  Electronic Control of Coupling}},\ }\href
  {https://onlinelibrary.wiley.com/doi/10.1002/qute.202200049} {\bibfield
  {journal} {\bibinfo  {journal} {Advanced Quantum Technologies}\ }\textbf
  {\bibinfo {volume} {5}},\ \bibinfo {pages} {2200049} (\bibinfo {year}
  {2022})}\BibitemShut {NoStop}%
\bibitem [{\citenamefont {Fox}()}]{Fox1970}%
  \BibitemOpen
  \bibfield  {author} {\bibinfo {author} {\bibfnamefont {M.}~\bibnamefont
  {Fox}},\ }\href@noop {} {\emph {\bibinfo {title} {Quantum Optics: An
  Introduction (Oxford University Press)}}},\ \bibinfo {number}
  {2006}\BibitemShut {NoStop}%
\bibitem [{\citenamefont {Stievater}\ \emph {et~al.}(2001)\citenamefont
  {Stievater}, \citenamefont {Li}, \citenamefont {Steel}, \citenamefont
  {Gammon}, \citenamefont {Katzer}, \citenamefont {Park}, \citenamefont
  {Piermarocchi},\ and\ \citenamefont {Sham}}]{Stievater2001}%
  \BibitemOpen
  \bibfield  {author} {\bibinfo {author} {\bibfnamefont {T.~H.}\ \bibnamefont
  {Stievater}}, \bibinfo {author} {\bibfnamefont {X.}~\bibnamefont {Li}},
  \bibinfo {author} {\bibfnamefont {D.~G.}\ \bibnamefont {Steel}}, \bibinfo
  {author} {\bibfnamefont {D.}~\bibnamefont {Gammon}}, \bibinfo {author}
  {\bibfnamefont {D.~S.}\ \bibnamefont {Katzer}}, \bibinfo {author}
  {\bibfnamefont {D.}~\bibnamefont {Park}}, \bibinfo {author} {\bibfnamefont
  {C.}~\bibnamefont {Piermarocchi}},\ and\ \bibinfo {author} {\bibfnamefont
  {L.~J.}\ \bibnamefont {Sham}},\ }\bibfield  {title} {\bibinfo {title} {{Rabi
  Oscillations of Excitons in Single Quantum Dots}},\ }\href@noop {} {\bibfield
   {journal} {\bibinfo  {journal} {Physical Review Letters}\ }\textbf {\bibinfo
  {volume} {87}},\ \bibinfo {pages} {133603} (\bibinfo {year}
  {2001})}\BibitemShut {NoStop}%
\bibitem [{\citenamefont {Kuhlmann}\ \emph {et~al.}(2013)\citenamefont
  {Kuhlmann}, \citenamefont {Houel}, \citenamefont {Brunner}, \citenamefont
  {Ludwig}, \citenamefont {Reuter}, \citenamefont {Wieck},\ and\ \citenamefont
  {Warburton}}]{Kuhlmann2013}%
  \BibitemOpen
  \bibfield  {author} {\bibinfo {author} {\bibfnamefont {A.~V.}\ \bibnamefont
  {Kuhlmann}}, \bibinfo {author} {\bibfnamefont {J.}~\bibnamefont {Houel}},
  \bibinfo {author} {\bibfnamefont {D.}~\bibnamefont {Brunner}}, \bibinfo
  {author} {\bibfnamefont {A.}~\bibnamefont {Ludwig}}, \bibinfo {author}
  {\bibfnamefont {D.}~\bibnamefont {Reuter}}, \bibinfo {author} {\bibfnamefont
  {A.~D.}\ \bibnamefont {Wieck}},\ and\ \bibinfo {author} {\bibfnamefont
  {R.~J.}\ \bibnamefont {Warburton}},\ }\bibfield  {title} {\bibinfo {title}
  {{A dark-field microscope for background-free detection of resonance
  fluorescence from single semiconductor quantum dots operating in a
  set-and-forget mode}},\ }\href@noop {} {\bibfield  {journal} {\bibinfo
  {journal} {Review of Scientific Instruments}\ }\textbf {\bibinfo {volume}
  {84}},\ \bibinfo {pages} {073905} (\bibinfo {year} {2013})}\BibitemShut
  {NoStop}%
\bibitem [{\citenamefont {Nakaoka}\ \emph {et~al.}(2006)\citenamefont
  {Nakaoka}, \citenamefont {Clark}, \citenamefont {Krenner}, \citenamefont
  {Sabathil}, \citenamefont {Bichler}, \citenamefont {Arakawa}, \citenamefont
  {Abstreiter},\ and\ \citenamefont {Finley}}]{Nakaoka2006}%
  \BibitemOpen
  \bibfield  {author} {\bibinfo {author} {\bibfnamefont {T.}~\bibnamefont
  {Nakaoka}}, \bibinfo {author} {\bibfnamefont {E.~C.}\ \bibnamefont {Clark}},
  \bibinfo {author} {\bibfnamefont {H.~J.}\ \bibnamefont {Krenner}}, \bibinfo
  {author} {\bibfnamefont {M.}~\bibnamefont {Sabathil}}, \bibinfo {author}
  {\bibfnamefont {M.}~\bibnamefont {Bichler}}, \bibinfo {author} {\bibfnamefont
  {Y.}~\bibnamefont {Arakawa}}, \bibinfo {author} {\bibfnamefont
  {G.}~\bibnamefont {Abstreiter}},\ and\ \bibinfo {author} {\bibfnamefont
  {J.~J.}\ \bibnamefont {Finley}},\ }\bibfield  {title} {\bibinfo {title}
  {{Direct observation of acoustic phonon mediated relaxation between coupled
  exciton states in a single quantum dot molecule}},\ }\href@noop {} {\bibfield
   {journal} {\bibinfo  {journal} {Physical Review B - Condensed Matter and
  Materials Physics}\ }\textbf {\bibinfo {volume} {74}} (\bibinfo {year}
  {2006})}\BibitemShut {NoStop}%
\bibitem [{\citenamefont {F{\"{o}}rstner}\ \emph {et~al.}(2003)\citenamefont
  {F{\"{o}}rstner}, \citenamefont {Weber}, \citenamefont {Danckwerts},\ and\
  \citenamefont {Knorr}}]{Forstner2003}%
  \BibitemOpen
  \bibfield  {author} {\bibinfo {author} {\bibfnamefont {J.}~\bibnamefont
  {F{\"{o}}rstner}}, \bibinfo {author} {\bibfnamefont {C.}~\bibnamefont
  {Weber}}, \bibinfo {author} {\bibfnamefont {J.}~\bibnamefont {Danckwerts}},\
  and\ \bibinfo {author} {\bibfnamefont {A.}~\bibnamefont {Knorr}},\ }\bibfield
   {title} {\bibinfo {title} {{Phonon-assisted damping of rabi oscillations in
  semiconductor quantum dots}},\ }\href
  {https://journals.aps.org/prl/abstract/10.1103/PhysRevLett.91.127401}
  {\bibfield  {journal} {\bibinfo  {journal} {Physical Review Letters}\
  }\textbf {\bibinfo {volume} {91}},\ \bibinfo {pages} {127401} (\bibinfo
  {year} {2003})}\BibitemShut {NoStop}%
\bibitem [{\citenamefont {Kaldewey}\ \emph {et~al.}(2017)\citenamefont
  {Kaldewey}, \citenamefont {L{\"{u}}ker}, \citenamefont {Kuhlmann},
  \citenamefont {Valentin}, \citenamefont {Chauveau}, \citenamefont {Ludwig},
  \citenamefont {Wieck}, \citenamefont {Reiter}, \citenamefont {Kuhn},\ and\
  \citenamefont {Warburton}}]{Kaldewey2017}%
  \BibitemOpen
  \bibfield  {author} {\bibinfo {author} {\bibfnamefont {T.}~\bibnamefont
  {Kaldewey}}, \bibinfo {author} {\bibfnamefont {S.}~\bibnamefont
  {L{\"{u}}ker}}, \bibinfo {author} {\bibfnamefont {A.~V.}\ \bibnamefont
  {Kuhlmann}}, \bibinfo {author} {\bibfnamefont {S.~R.}\ \bibnamefont
  {Valentin}}, \bibinfo {author} {\bibfnamefont {J.~M.}\ \bibnamefont
  {Chauveau}}, \bibinfo {author} {\bibfnamefont {A.}~\bibnamefont {Ludwig}},
  \bibinfo {author} {\bibfnamefont {A.~D.}\ \bibnamefont {Wieck}}, \bibinfo
  {author} {\bibfnamefont {D.~E.}\ \bibnamefont {Reiter}}, \bibinfo {author}
  {\bibfnamefont {T.}~\bibnamefont {Kuhn}},\ and\ \bibinfo {author}
  {\bibfnamefont {R.~J.}\ \bibnamefont {Warburton}},\ }\bibfield  {title}
  {\bibinfo {title} {{Demonstrating the decoupling regime of the
  electron-phonon interaction in a quantum dot using chirped optical
  excitation}},\ }\href
  {https://journals.aps.org/prb/abstract/10.1103/PhysRevB.95.241306} {\bibfield
   {journal} {\bibinfo  {journal} {Physical Review B}\ }\textbf {\bibinfo
  {volume} {95}},\ \bibinfo {pages} {241306} (\bibinfo {year}
  {2017})}\BibitemShut {NoStop}%
\bibitem [{\citenamefont {Schall}\ \emph {et~al.}(2021)\citenamefont {Schall},
  \citenamefont {Deconinck}, \citenamefont {Bart}, \citenamefont {Florian},
  \citenamefont {Helversen}, \citenamefont {Dangel}, \citenamefont {Schmidt},
  \citenamefont {Bremer}, \citenamefont {Bopp}, \citenamefont {H{\"{u}}llen},
  \citenamefont {Gies}, \citenamefont {Reuter}, \citenamefont {Wieck},
  \citenamefont {Rodt}, \citenamefont {Finley}, \citenamefont {Jahnke},
  \citenamefont {Ludwig},\ and\ \citenamefont {Reitzenstein}}]{Schall2021}%
  \BibitemOpen
  \bibfield  {author} {\bibinfo {author} {\bibfnamefont {J.}~\bibnamefont
  {Schall}}, \bibinfo {author} {\bibfnamefont {M.}~\bibnamefont {Deconinck}},
  \bibinfo {author} {\bibfnamefont {N.}~\bibnamefont {Bart}}, \bibinfo {author}
  {\bibfnamefont {M.}~\bibnamefont {Florian}}, \bibinfo {author} {\bibfnamefont
  {M.}~\bibnamefont {Helversen}}, \bibinfo {author} {\bibfnamefont
  {C.}~\bibnamefont {Dangel}}, \bibinfo {author} {\bibfnamefont
  {R.}~\bibnamefont {Schmidt}}, \bibinfo {author} {\bibfnamefont
  {L.}~\bibnamefont {Bremer}}, \bibinfo {author} {\bibfnamefont
  {F.}~\bibnamefont {Bopp}}, \bibinfo {author} {\bibfnamefont {I.}~\bibnamefont
  {H{\"{u}}llen}}, \bibinfo {author} {\bibfnamefont {C.}~\bibnamefont {Gies}},
  \bibinfo {author} {\bibfnamefont {D.}~\bibnamefont {Reuter}}, \bibinfo
  {author} {\bibfnamefont {A.~D.}\ \bibnamefont {Wieck}}, \bibinfo {author}
  {\bibfnamefont {S.}~\bibnamefont {Rodt}}, \bibinfo {author} {\bibfnamefont
  {J.~J.}\ \bibnamefont {Finley}}, \bibinfo {author} {\bibfnamefont
  {F.}~\bibnamefont {Jahnke}}, \bibinfo {author} {\bibfnamefont
  {A.}~\bibnamefont {Ludwig}},\ and\ \bibinfo {author} {\bibfnamefont
  {S.}~\bibnamefont {Reitzenstein}},\ }\bibfield  {title} {\bibinfo {title}
  {{Bright Electrically Controllable Quantum‐Dot‐Molecule Devices
  Fabricated by In Situ Electron‐Beam Lithography}},\ }\href
  {https://onlinelibrary.wiley.com/doi/10.1002/qute.202100002} {\bibfield
  {journal} {\bibinfo  {journal} {Advanced Quantum Technologies}\ }\textbf
  {\bibinfo {volume} {4}},\ \bibinfo {pages} {2100002} (\bibinfo {year}
  {2021})}\BibitemShut {NoStop}%
\bibitem [{\citenamefont {Bouarissa}\ and\ \citenamefont
  {Aourag}(1999)}]{Bouarissa1999}%
  \BibitemOpen
  \bibfield  {author} {\bibinfo {author} {\bibfnamefont {N.}~\bibnamefont
  {Bouarissa}}\ and\ \bibinfo {author} {\bibfnamefont {H.}~\bibnamefont
  {Aourag}},\ }\bibfield  {title} {\bibinfo {title} {{Effective masses of
  electrons and heavy holes in InAs, InSb, GaSb, GaAs and some of their ternary
  compounds}},\ }\href
  {https://doi.org/https://doi.org/10.1016/S1350-4495(99)00020-1} {\bibfield
  {journal} {\bibinfo  {journal} {Infrared Physics \& Technology}\ }\textbf
  {\bibinfo {volume} {40}},\ \bibinfo {pages} {343} (\bibinfo {year}
  {1999})}\BibitemShut {NoStop}%
\end{thebibliography}%

\end{document}